\documentclass[letterpaper, 10 pt, conference]{ieeeconf}  % Comment this line out if you need a4paper

\IEEEoverridecommandlockouts                              % This command is only needed if 
\usepackage{cite}
\usepackage{amsmath,amssymb,amsfonts}
\usepackage{algorithmic}
\usepackage{graphicx}
\usepackage{textcomp}
\usepackage{xcolor}
\usepackage{tikz}
\usepackage{tikzscale}
\usepackage{pgfplots}
\usepackage{eso-pic}
\usepackage{pgfornament}
\pgfplotsset{compat=newest,
	every tick label/.append style={scale=0.7},
	every axis/.append style={
		label style={scale=0.8}},
}
\usetikzlibrary{plotmarks}
\definecolor{black}{rgb}{0,0,0}
\definecolor{matblue1}{rgb}{0,0.4470,0.7410}
\definecolor{matred1}{rgb}{0.85,0.325,0.098}
\definecolor{matyel1}{rgb}{0.9290,0.6940,0.1250}
\definecolor{silver}{rgb}{0.7529412, 0.7529412, 0.7529412}
\definecolor{fillgray}{rgb}{0.9,0.9,0.9}
\definecolor{purple}{rgb}{0.4940, 0.1840, 0.5560}
\definecolor{darkred}{rgb}{0.6350, 0.0780, 0.1840}
\definecolor{gray}{rgb}{0.65,0.65,0.65}

\newcommand{\blueline}{\raisebox{2pt}{\tikz{\draw[-,matblue1,solid,line width = 0.9pt](0,0) -- (3mm,0);}}}
\newcommand{\redline}{\raisebox{2pt}{\tikz{\draw[-,matred1,dashed,line width = 0.9pt](0,0) -- (3mm,0);}}}
\newcommand{\yelline}{\raisebox{2pt}{\tikz{\draw[-,matyel1,dash dot,line width = 0.9pt](0,0) -- (3mm,0);}}}
\newcommand{\grayline}{\raisebox{2pt}{\tikz{\draw[-,gray,dotted,line width = 0.9pt](0,0) -- (3mm,0);}}}

\newcommand{\bluedashsquare}{
	\begin{tikzpicture}[baseline=-0.54ex]
		\draw[-,matblue1,solid,line width=0.9pt](0,-0.12) -- (0.24,-0.12) -- (0.24,0.12) -- (0,0.12) -- cycle;
		\draw[-,matblue1,dashed,line width = 0.9pt](-0.2,0) -- (0.44,0);
\end{tikzpicture}}

\newcommand{\bluedashtriangle}{
	\begin{tikzpicture}[baseline=-0.54ex]
		\draw[-,matblue1,solid,line width=0.9pt](-0.07,-0.12) -- (0.05,0.12) -- (0.17,-0.12) -- cycle;
		\draw[-,matblue1,dashed,line width = 0.9pt](-0.2,0) -- (0.3,0);
\end{tikzpicture}}

\newcommand{\redcircle}{\begin{tikzpicture}[baseline=0.35ex]
		\draw[-,matred1,solid,line width=0.9pt] (0.12,0.12) circle [radius=0.12];
		\draw[-,matred1,dashed,line width = 0.9pt](-0.2,0.12) -- (0.44,0.12);
\end{tikzpicture}}

\newcommand{\redplus}{\tikz{\node[matred1,mark size=0.7ex]{\pgfuseplotmark{+}};}}

\newcommand{\varianceband}{\begin{tikzpicture}
		\filldraw[fill=fillgray, draw=black] (0,0) rectangle (1ex,1ex);
\end{tikzpicture}}

\newcommand{\silverdashcircle}{\begin{tikzpicture}[baseline=0.4ex]
		\draw[-,silver,solid,line width=0.9pt] (0.12,0.12) circle [radius=0.12];
		\draw[-,silver,dashed,line width = 0.9pt](-0.2,0.12) -- (0.44,0.12);
\end{tikzpicture}}
\newcommand{\fillred}{\tikz{\draw[fill=matred1,draw=black](0.1,0.1) rectangle (0.25,0.25);}}
\newcommand{\fillblue}{\tikz{\draw[fill=matblue1,draw=black](0.1,0.1) rectangle (0.25,0.25);}}
\newcommand{\fillyel}{\tikz{\draw[fill=matyel1,draw=black](0.1,0.1) rectangle (0.25,0.25);}}
\newcommand{\dashred}{\raisebox{2pt}{\tikz{\draw[-,matred1,dashed,line width = 0.9pt](0,0) -- (3mm,0);}}}

\newcommand{\purptriangle}{\raisebox{-0.8pt}{\tikz{\node[purple,mark size=0.7ex]{\pgfuseplotmark{triangle*}};}}}
\makeatletter
\providecommand\add@text{}
\newcommand\tagaddtext[1]{%
	\gdef\add@text{#1\gdef\add@text{}}}% 
\renewcommand\tagform@[1]{%
	\maketag@@@{\llap{\add@text\quad}(\ignorespaces#1\unskip\@@italiccorr)}%
}
\makeatother
\newcounter{definition}

\newcounter{assumption}

\newcounter{example}

\title{\LARGE \bf
Position-Dependent Snap Feedforward: A Gaussian Process Framework
}

\author{Max van Haren$^{1}$, Maurice Poot$^{1}$, Jim Portegies$^{2}$ and Tom Oomen$^{1,3}$% <-this % stops a space
\thanks{This work is part of the research programme VIDI with project number 15698, which is (partly) financed by the Netherlands Organisation for Scientific Research (NWO). In addition, this research has received funding from the ECSEL Joint Undertaking under grant agreement 101007311 (IMOCO4.E). The Joint Undertaking receives support from the European Union’s Horizon 2020 research and innovation programme.}% <-this % stops a space
\thanks{$^{1}$Max van Haren, Maurice Poot and Tom Oomen are with the Control Systems Technology Section, Department of Mechanical Engineering, Eindhoven University of Technology, Eindhoven, The Netherlands.
        {\tt\small m.j.v.haren@tue.nl}}%
\thanks{$^{2}$Jim Portegies is with the Centre for Analysis, Scientific Computing and Applications, Department of Mathematics and Computer Science, Eindhoven University of Technology, Eindhoven, The Netherlands.}
\thanks{$^{3}$Tom Oomen is with the Delft Center for Systems and Control, Delft University of Technology, Delft, The Netherlands.}}

\AddToShipoutPictureBG*{%
	\AtPageUpperLeft{%
		\setlength\unitlength{1in}%
		\hspace*{\dimexpr0.5\paperwidth\relax}%% change \dimexpr0.5\paperwidth\relax appropriately
		\makebox(0,-0.75)[c]{\begin{tabular}{c c}
				Max van Haren, Position-Dependent Snap Feedforward: A Gaussian Process Framework, \\
				To appear in {\em 2022 IEEE American Control Conference}, Atlanta, Georgia, 2022%
		\end{tabular}}
}}

\begin{document}

\maketitle
\thispagestyle{empty}
\pagestyle{empty}
\begin{abstract}
	Mechatronic systems have increasingly high performance requirements for motion control. The low-frequency contribution of the flexible dynamics, i.e. the compliance, should be compensated for by means of snap feedforward to achieve high accuracy. Position-dependent compliance, which often occurs in motion systems, requires the snap feedforward parameter to be modeled as a function of position. Position-dependent compliance is compensated for by using a Gaussian process to model the snap feedforward parameter as a continuous function of position. A simulation of a flexible beam shows that a significant performance increase is achieved when using the Gaussian process snap feedforward parameter to compensate for position-dependent compliance.
\end{abstract}

%%%%%%%%%%%%%%%%%%%%%%%%%%%%%%%%%%%%%%%%%%%%%%%%%%%%%%%%%%%%%%%%%%%%%%%%%%%%%%%%
%-------------------------------------------sections---------------------------------------------------
%%%%%%%%%%%%%%%%%%%%%%%%%%%%%%%%%%%%%%%%%%%%%%%%%intro%%%%%%%%%%%%%%%%%%%%%%%%%%%%%%%%%%%%%%%%%%%%%%%%%%
\section{Introduction}
%intro general
Feedforward control is fundamental for the tracking performance of motion systems, including wafer stages \cite{deRozario2017b} and printing systems \cite{Bolder2014}. Traditionally, feedforward control is based on manual tuning. Recently, due to the increase of computational power, the focus has shifted towards learning feedforward from data \cite{Oomen2020}. For example, Iterative Learning Control (ILC) achieves high tracking performance by learning feedforward in a trial-to-trial fashion \cite{Bristow2006}. In contrast, fast motion is realised by designing lightweight systems, introducing dominant flexible dynamics affecting the tracking performance \cite{Ronde2014}. In addition, the combination of moving bodies and flexible dynamics introduce position-dependent behavior \cite{deSilva2007,Voorhoeve2021}. \par 
% acc feedforward rigid body dynamics
Classical Linear Time-Invariant (LTI) acceleration feedforward compensates for the rigid-body dynamics of a system. In this case, the feedforward signal is scaled to the acceleration of the reference trajectory \cite{Oomen2019}. However, a well-tuned acceleration feedforward does not compensate for flexible dynamics \cite{Boerlage2004}. \par
% snap
Flexible dynamics lead to a situation where the compliance, i.e. the low-frequency contribution of the flexible dynamics, is compensated for by means of snap feedforward \cite{Boerlage2004}. Snap feedforward uses the scaled fourth derivative of the reference. The snap feedforward parameter can be tuned online and improves performance for both Single-Input, Single-Output (SISO) systems \cite{Boerlage2004} and multiple-input, multiple-output systems \cite{Boerlage2006}.\par
% pos dep (snap)?
Many systems contain position-dependent behavior that introduces position-dependent compliance \cite{Voorhoeve2016,deRozario2021,Kontaras2016}, which necessitates the need for a position-dependent compensation thereof. For this purpose, the snap feedforward parameter can be determined in a grid and estimated with, e.g., spline or linear interpolation. However, these interpolations have approximation errors since the dependency between position and snap feedforward parameters is generally unknown. \par
% LPV control
High motion control performance for systems with position-dependent dynamics can furthermore be achieved through the use of Linear Parameter Varying (LPV) control of the system. First, LTI dynamics can be scheduled according to the current configuration of the LPV system, resulting in high control performance for e.g. wafer stages \cite{deRozario2017b,Wassink2005} or xy-positioning tables \cite{Toth2011}. Second, ILC can be extended for LPV systems, which results in high performance through learning \cite{deRozario2017}. LPV model-free approaches are investigated in \cite{Formentin2016}, directly learning LPV controllers from data, but are at present not competitive with model-based designs. Therefore, LPV control requires LPV modeling, which is often very challenging and the high modeling cost and complexity are usually not justified for industrial control applications.\par
%ALTHOUGH
Although feedforward design has improved significantly with respect to traditional acceleration feedforward, a snap feedforward with systematic tuning for position-dependent compliance, capable of estimation at any arbitrary position, is currently lacking. This paper models the snap feedforward parameter as a continuous function of position by means of a Gaussian Process (GP) \cite{Rasmussen2004,Pillonetto2014}, which allows for the compensation of position-dependent compliance without an LPV model. In addition, a GP is non-parametric and therefore does not require an assumption on the parametric form between the position and the snap feedforward parameter. In this paper, the feedforward parameters of a system are learned in a trial-to-trial fashion using ILC with Basis Functions (ILCBF) \cite{vandeWijdeven2010}. The contributions include:
\begin{enumerate}
	\item[C1] a framework to model the snap feedforward parameter as a function of position with a GP (Section~\ref{sec:GPSnap}),
	\item[C2] ILCBF to automatically learn the snap feedforward parameter, which is directly used in the GP (Section~\ref{sec:ILCBF}),
	\item[C3] application to a benchmark example, confirming the capabilities of the framework (Section~\ref{sec:results}).
\end{enumerate}
\hfill\par
\textbf{Notation:} Systems are SISO and discrete-time, unless stated otherwise. Continuous time systems are transformed in their discrete-time counterpart using finite difference approximation. The trial number is indicated with the index $j$. Signals are assumed to be of length $N$. The weighted 2-norm of a vector $x \in \mathbb{R}^N$ is denoted as $\| x \|_W := \sqrt(x^\top W x)$, where $W \in \mathbb{R}^{N\times N}$ is a weighting matrix. Matrix $A\in\mathbb{R}^{N\times N}$ is positive (semi-)definite if and only if $x^\top A x \geq 0, \; \forall x \neq 0 \in \mathbb{R}^N$ and is denoted as $A\succeq 0$.
%%%%%%%%%%%%%%%%%%%%%%%%%%%%%%%%%%%%%%%%%%%%Problem formulation%%%%%%%%%%%%%%%%%%%%%%%%%%%%%%%%%%%%%%%%%%%
\section{Problem Formulation}
\label{sec:problemFormulation}
In this section, the problem for determining a feedforward controller for position-dependent flexible dynamics is formulated. First, a general description for system with position-dependent flexible dynamics is given. Second, feedforward design, including acceleration and snap feedforward, is investigated. Finally, the problem addressed in this paper is defined.
%------------------------------------------------------------------------------------------------------------
\subsection{Considered Class of Position-Dependent Systems}
\begin{figure}[tbp]
	\centerline{\includegraphics[width=\columnwidth]{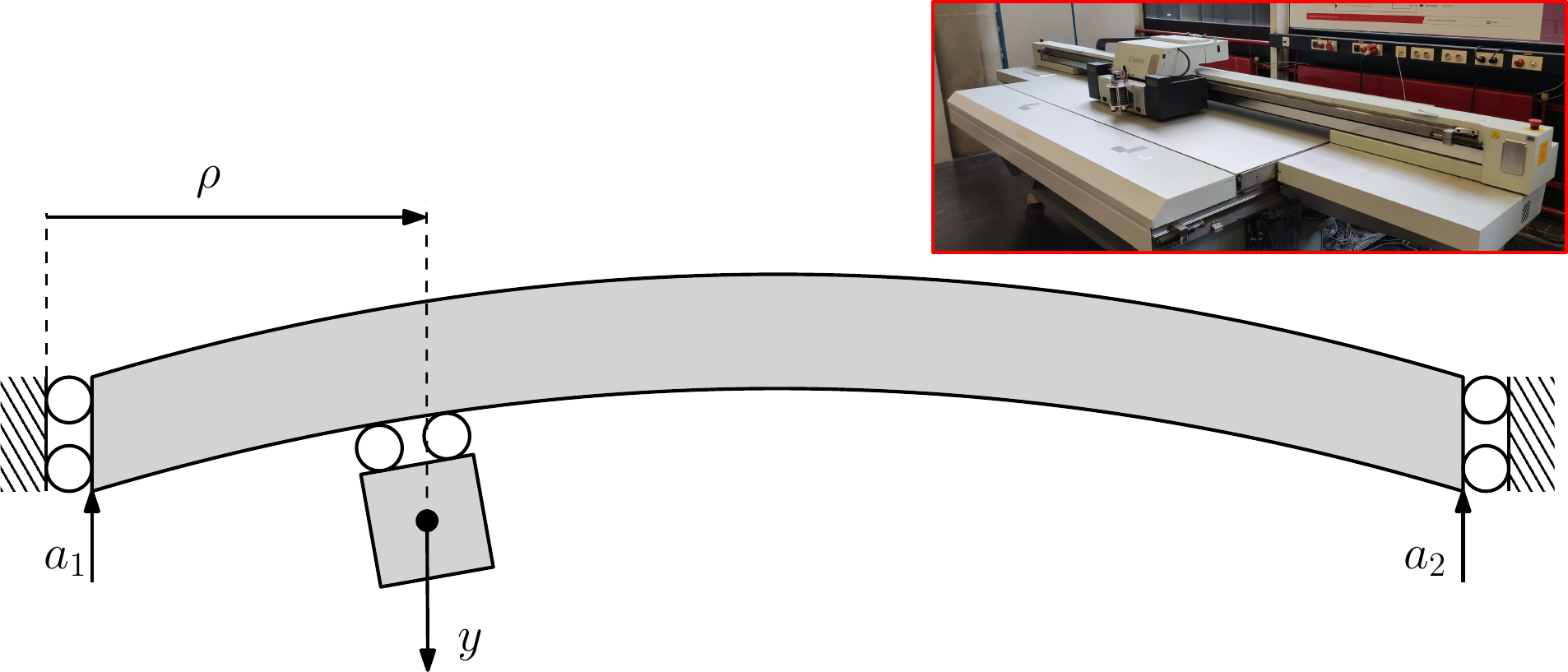}}
	\caption{Benchmark system for position-dependent snap feedforward, representing an H-type gantry commonly used in e.g. semiconductor back-end machines or large-format printing systems seen in the top right. The first mode shape of the flexible beam is displayed, with two actuators $a_1$ and $a_2$ and the output $y$. The performance location can vary, indicated by the scheduling variable $\rho$.}
	\label{fig:beamFigure}
\end{figure}

The considered class of systems with position-dependent flexible modes are encompassed using spatially distributed LTI systems \cite[Section~3.2]{wodek2004,Moheimani2003}
\begin{equation}
	\label{eq:LPVTF}
	\begin{split}
		y(k) &= G(\rho,q^{-1})u(k), \\
		G(\rho,q^{-1}) &= \underbrace{\sum_{l=1}^{n_{RB}}G_{l,RB}(q^{-1})}_{\text{rigid-body modes}}+\underbrace{\sum_{i=1}^{n_f}D_i(\rho)G_{i,f}(q^{-1})}_{\text{flexible modes}},
	\end{split}
\end{equation}
with scheduling variable $\rho\in\mathcal{D}$, where $\mathcal{D}$ is the parameter space and $q$ the shift operator, i.e., $q^{-\tau}s(k)=s(k-\tau)$. Note that $\rho$ is not limited to the position, but can be used as different scheduling variable, see e.g. \cite[Fig.~7]{Abramovitch2015}. The mode shapes are represented by $D_i(\rho)=c_i(\rho)b_i^\top(\rho) \in \mathbb{R}$ and $n_{RB},\, n_{f}\in\mathbb{N}_+$ are the amount of rigid-body and flexible modes. The systems $G_{l,RB}$ and $G_{i,f}$ are described as \cite{wodek2004}
\begin{equation}
	\begin{split}
		G_{l,RB}(q^{-1}) &= \frac{c_lb_l^\top T_s^2}{(1-q^{-1})^2}, \\
		G_{i,f}(q^{-1}) &= \frac{1}{\frac{(1-q^{-1})^2}{T_s^2}+2\zeta_i\omega_i\frac{1-q^{-1}}{T_s}+\omega_i^2},
	\end{split}
\end{equation}
with $T_s$ the sampling time and $\zeta_i,\omega_i \in\mathbb{R}_+$ the mode damping and mode frequency, respectively. A benchmark example is the flexible beam with varying performance location, as seen in Fig.~\ref{fig:beamFigure}. \par
\begin{figure}[tbp]
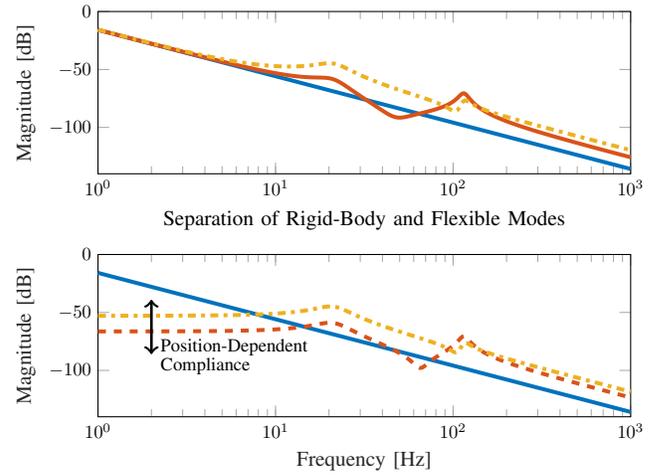

	\centering
	\input{beamBodes}
	\input{bodesSeperation}
	\caption{Top: bode magnitude diagram of the frozen transfer function $G(\rho,q^{-1})$ evaluated at $\rho=0$ mm (\protect\blueline), $\rho=35$ mm (\protect\redline) and $\rho=250$ mm (\protect\yelline). Bottom: the separation of rigid-body (\protect\blueline) and flexible modes at $\rho=35$ mm (\protect\redline) and $\rho=250$ mm (\protect\yelline) for the flexible beam in Fig.~\ref{fig:beamFigure}.}
	\label{fig:beamBodes}
\end{figure}

For a fixed value of $\rho$, the system $G(\rho,q^{-1})$ is an LTI system, which is called the frozen dynamics of the system. Several bode magnitude diagrams evaluated at different values of $\rho$ can be seen in Fig.~\ref{fig:beamBodes}
%----------------------------------------------------------------------------------------------------------
\subsection{Feedforward Design and the Limitation of Acceleration Feedforward}
\begin{figure}[tbp]
	\centerline{\includegraphics[width=0.96\columnwidth]{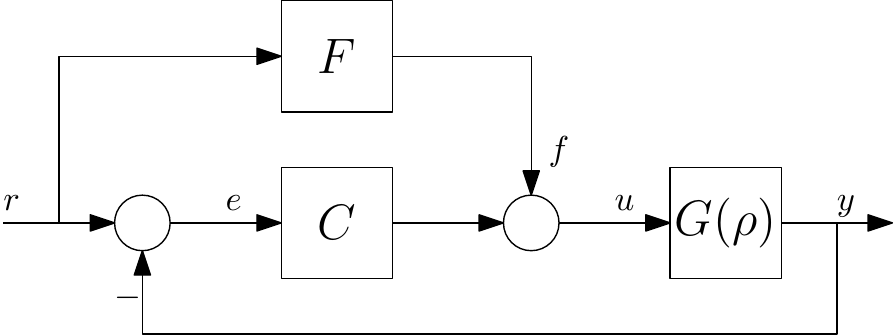}}
	\caption{Control structure considered in this paper, consisting of feedforward controller $F$, feedback controller $C$ and the system $G(\rho)$.}
	\label{fig:controlLayout}
\end{figure}

The main goal of feedforward is to minimize the error $e$, given a reference signal $r$, see Fig.~\ref{fig:controlLayout}. 
Consider the continuous-time LTI position-independent system equivalent to \eqref{eq:LPVTF},
\begin{equation}
	\label{eq:mechanics}
	G_0(s)=\sum_{l=1}^{n_{RB}} \frac{c_{l} b_{l}^{\top}}{s^{2}}+\sum_{i=1}^{n_{f}} \frac{c_{i} b_{i}^{\top}}{s^{2}+2 \zeta_{i} \omega_{i} s+\omega_{i}^{2}},
\end{equation}
where $s$ is the continuous-time indeterminate. The objective of feedforward, for the plant $G_0$, is to minimize
\begin{equation*}
	\label{eq:error}
	e(s) = S(s)r(s)-S(s)G_0(s)F(s)r(s),
\end{equation*}
where
\begin{equation*}
	\label{eq:sensitivity}
	S(s) = (I+G_0(s)C(s))^{-1}.
\end{equation*}
Typical feedforward, for systems with rigid-body modes which are not suspended, consists of acceleration feedforward,
\begin{equation}
	\label{eq:accelerationFeedforward}
	{F}_a(s) = \hat{m}s^2,
\end{equation}
with $\hat{m}$ an estimate of the mass of the system. Consider the open loop tracking error when assuming the system $G_0$ has only one translational rigid-body mode and well-tuned acceleration feedforward, i.e., $\sum_{i=l}^{n_{RB}}c_lb_l^\top=\frac{1}{m}$ in \eqref{eq:mechanics} and $\hat{m}=m$,
\begin{equation}
	\label{eq:accerror}
	\small
	\begin{split}
		e_o(s) &= r(s)-G_0(s)F_a(s)r(s) \\
		&= -\sum_{i=1}^{n_f}\frac{c_i\,b_i^\top}{s^2-2\omega_i\zeta_is+\omega_i^2}ms^2r(s).
	\end{split}
\end{equation}
For low frequencies, the relation between the error in \eqref{eq:accerror} and the acceleration of the reference is
\begin{equation}
	\label{eq:complianceError}
	\lim_{s\rightarrow0}\bigg(\frac{e_o(s)}{s^2{r}(s)}\bigg) = -m\sum_{i=1}^{n_f}\frac{ c_{i} b_i^\top}{\omega_i^2},
\end{equation}
which shows that the open loop servo error for low frequencies remains when using acceleration feedforward due to the contribution of the compliance.
%----------------------------------------------------------------------------------------
\subsection{Snap Feedforward}
Flexible systems necessitate a compensation of the flexible dynamics during constant acceleration, as seen in \eqref{eq:complianceError}, which is explicitly compensated for by means of snap feedforward. For this purpose, the acceleration feedforward in \eqref{eq:accelerationFeedforward} is extended with additional snap feedforward $\delta s^4$, leading to
\begin{equation}
	\label{eq:snap1}
	{F}_s(s) = {m}s^2 +\delta s^4,
\end{equation}
with $\delta$ the snap feedforward parameter. The open loop servo error in \eqref{eq:accerror} is equal to zero by designing $\delta$ as,
\begin{equation*}
	G_0^{-1}(s)\frac{1}{s^4}-{m}\frac{1}{s^2}.
\end{equation*}
Again assuming one translational rigid body mode, the low-frequency contribution of this controller is equal to
\begin{equation*}
	\delta = \lim_{s\rightarrow 0}\bigg( {G}_0^{-1}(s)\frac{1}{s^4}-{m}\frac{1}{s^2}\bigg) = -{m}^2\sum_{i=1}^{n_f}\frac{ c_{i} b_i^\top}{\omega_i^2}.
\end{equation*}
Hence, the feedforward controller in \eqref{eq:snap1} results in
\begin{equation}
	{F}_s(s) = {m}s^2-{m}^2\sum_{i=1}^{n_f}\frac{ c_{i} b_i^\top}{\omega_i^2}\; s^4.
\end{equation}
Note that for position-independent systems, the value of $\delta$ is a real constant and can be learned or tuned online.
%----------------------------------------------------------------------------------------

\subsection{Problem Definition}
Position-independent snap feedforward does not compensate for position-dependent compliance of the system $G(\rho,q^{-1})$. LPV modeling of the position-dependent compliance is often not feasible and a parametric form between the position and the snap feedforward parameter, that is used in interpolation, is generally unknown. \par 
Hence, the aim of this paper is modeling the snap feedforward parameter $\delta$ in \eqref{eq:snap1} as a continuous function of position, i.e. $\delta(\rho)$, such that it compensates for position-dependent compliance, without an LPV model or parametric form.
%%%%%%%%%%%%%%%%%%%%%%%%%%%%%%%%%%%%%%%%%%%%%%%%%GP Snap%%%%%%%%%%%%%%%%%%%%%%%%%%%%%%%%%%%%%%%%%%%%%%%%%%
\section{Position-dependent Snap Feedforward using Gaussian processes}
\label{sec:GPSnap}
In this section, the snap feedforward parameter is modeled as a function of position, i.e., $\delta(\rho)$, such that it can compensate for position-dependent compliance, hence constituting contribution C1. First, GPs are investigated, including the covariance function, the prior distribution and the posterior distribution. Finally, a method to model the snap feedforward parameter as a function of position by means of a GP is shown.
%---------------------------------------------------------
\subsection{Gaussian Processes}
A GP is defined as a collection of random variables $f(\rho)$, indexed by $\rho \in \mathbb{R}^{\mathcal{D}}$, such that the joint distribution of any finite subset of random variables is multivariate Gaussian. A GP is written as
\begin{equation}
	\label{eq:GPDef}
	f(\rho) \sim \mathcal{GP}\big(m(\rho),k(\rho,\rho^\prime)\big),
\end{equation}
and is completely determined by the covariance function $k(\rho,\rho^\prime)$ and the mean function $m(\rho)$,
\begin{equation}
	\label{eq:meanAndCovFunction}
	\begin{aligned}
		k\left(\rho, \rho^{\prime}\right) &=\mathbb{E}\left[(f(\rho)-m(\rho))\left(f\left(\rho^{\prime}\right)-m\left(\rho^{\prime}\right)\right)\right], \\
		m(\rho) &=\mathbb{E}[f(\rho)].
	\end{aligned}
\end{equation}
The mean function $m$ can be interpreted as the mean at any input point and the covariance function $k$ as the similarity between values of $f(\rho)$ on different inputs $\rho$. The covariance function is discussed in Section~\ref{sec:covarianceFunction} and the mean function is assumed to be zero. Note that this is not necessary, see \cite[Section~2.7]{Rasmussen2004}. Training data for a GP are defined by sampling the function $f$ on inputs and measuring the (noisy) training outputs $y$
\begin{equation}
	\label{eq:functionSamples}
	\begin{aligned}
		y &=f(\rho)+\epsilon, \\
		\text { where } \epsilon & \sim \mathcal{N}\left(0, \sigma_{\epsilon}^{2} I\right),
	\end{aligned}
\end{equation}
and $\sigma_\epsilon^2$ is the variance of the noise acting on the output.
%--------------------------------------------------------
\subsection{Covariance Function}
\label{sec:covarianceFunction}
The covariance function or kernel specifies the covariance between the inputs $\rho$ and $\rho^\prime$. An example is the squared exponential or Radial Basis Function (RBF) covariance function,
\begin{equation}
	\label{eq:RBFKernel}
	k_{RBF}\left(\rho, \rho^{\prime}\right)=\sigma_{f}^{2} e^{-\frac{1}{2}\left(\rho-\rho^{\prime}\right)^{\top} \ell\left(\rho-\rho^{\prime}\right)},
\end{equation}
which shows that the entries for the covariance function are low when the inputs are far away from each other and close to $\sigma_f^2$ when they are close to each other. In addition, the $\sigma_f^2$ and $\ell$ are the so-called hyperparameters, respectively the signal variance and length scale, that are optimized based on data using marginal likelihood optimization \cite[Chapter~5]{Rasmussen2004}.
%---------------------------------------------------------
\subsection{Prior and Posterior Distribution}
\begin{figure}[tbp]
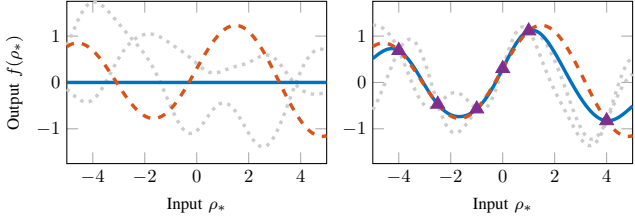

	\centering
	\input{prior}
	\input{posterior}
	\caption{Left: two samples (\protect\grayline) drawn from the prior, the prior mean (\protect\blueline) and a function to regress (\protect\dashred). Right: two samples (\protect\grayline) drawn from the posterior, the posterior mean (\protect\blueline), a function to regress (\protect\dashred) and the training data (\protect\purptriangle).}
	\label{fig:priorPosterior}
\end{figure}

In a GP, function predictions are made using the posterior distribution, which is the prior distribution conditioned on function observations. Consider a finite set of test inputs $P_* \in \mathbb{R}^{n_*\times \mathcal{D}}$, the prior distribution of a GP is formulated as
\begin{equation}
	f\left(P_*\right) \sim \mathcal{N}\left(0, K\left(P_*, P_*\right)\right),
\end{equation}
that is, evaluate the covariance function at $P_*$ and take the associated Gaussian distribution. Several samples drawn from this distribution can be seen in Fig.~\ref{fig:priorPosterior}. The test outputs $f\left(P_*\right)$ are jointly distributed with the training outputs as
\begin{equation}
	\label{eq:postDistr}
	\left[\begin{array}{c}
		y \\
		\!\!f\left(P_*\right)\!\!
	\end{array}\right] \sim \mathcal{N}\left(0,\left[\begin{array}{cc}
		\! K(P, P)+\sigma_{n}^{2}I\!\!\!     & K\left(P, P_*\right)\!\! \\
		K\left(P_*, P\right)\!                        & K\left(P_*, P_*\right)\!\!
	\end{array}\right]\right),
\end{equation}
with kernel matrices $K(P,P)\in\mathbb{R}^{n\times n}$, $K(P_*,P_*)\in\mathbb{R}^{n_*\times n_*}$ and $K(P,P_*)=K(P_*,P)^\top \in \mathbb{R}^{n\times n_*}$, with $n$ the amount of training outputs. The parameter $\sigma_n^2$ is an approximation of the noise acting on the output $\sigma_\epsilon^2$, seen in \eqref{eq:functionSamples}, and an additional hyperparameter. The training inputs $P \in \mathbb{R}^{n\times \mathcal{D}}$ and test inputs $P_*$ can be any single point or vector of positions, i.e.,
\begin{equation}
	\label{eq:GPinputs}
	\begin{split}
		P_* &= \begin{bmatrix} \rho_{*,1} & \rho_{*,2} & \cdots & \rho_{*,n_*}
		\end{bmatrix}^\top,\\
		P&=\begin{bmatrix}
			\rho_{1} & \rho_{2} & \cdots & \rho_{n}
		\end{bmatrix}^{\top}.
	\end{split}
\end{equation}
Using the joint distribution in \eqref{eq:postDistr} and Bayes' rule, estimations are made using the posterior distribution
\begin{equation}
	\label{eq:postEstimation}
	f\left(P_*\right) \Big|\left[P_*, P, y\right] \sim \mathcal{N}\left(\bar{f}\left(P_*\right), \operatorname{cov}\left(f\left(P_*\right)\right)\right),
\end{equation}
where
\begin{equation}
	\label{eq:GPpostmeanpostvar}
	\small
	\begin{aligned}
		\bar{f}\left(P_*\right)\equiv \; \mathbb{E}\left[f\left(P_*\right)\right]&=K_{*}^{\top} \big(K(P, P)+\sigma_{n}^{2}\big)^{-1} y, \\
		\operatorname{cov}\left(f\left(P_*\right)\right)&=K_{* *}-K_{*}^{\top} \big(K(P, P)+\sigma_{n}^{2}\big)^{-1} K_{*} .
	\end{aligned}
\end{equation}
The prior and posterior enable the user to train a GP with function observations and estimate function values.
%----------------------------------------------------------------------------------------------
\subsection{Gaussian Processes for Snap Feedforward}
In this section, the snap feedforward parameter is modeled as a continuous function of position to compensate for position-dependent compliance by means of a GP, i.e.,
\begin{equation}
	\label{eq:GPSnapParameter}
	\delta(\rho) := f(\rho) \sim \mathcal{GP}(m(\rho),k(\rho,\rho^\prime)).
\end{equation}
The training data for the GP, seen in \eqref{eq:functionSamples}, is defined as
\begin{equation}
	\label{eq:trainingData}
	y = \begin{bmatrix} 
		\delta_{1} & \cdots & \delta_{n}
	\end{bmatrix}^\top \in \mathbb{R}^{n\times 1},
\end{equation}
where $\delta_{i}$ is determined using the method in Section~\ref{sec:ILCBF}. The training inputs in \eqref{eq:GPinputs}, for systems with position-dependent flexible modes, are equal to the scheduling variables $\rho$ where the snap feedforward parameters $\delta_i$ are determined. \par
Snap feedforward parameters can be estimated on positions $\rho_*$ using
\begin{equation}
	\label{eq:estTheta}
	\delta(\rho_*) = K(P,\rho_*)^\top \big(K(P, P)+\sigma_{n}^{2}\big)^{-1} y,
\end{equation}
with $y$ from \eqref{eq:trainingData}. Using \eqref{eq:estTheta}, snap feedforward parameters can be estimated on unknown positions using the training data, such that position-dependent compliance can be compensated for. 
%%%%%%%%%%%%%%%%%%%%%%%%%%%%%%%%%%%%%%%%%%%%%%%%%ILCBF%%%%%%%%%%%%%%%%%%%%%%%%%%%%%%%%%%%%%%%%%%%%%%%%%%%%
\section{Learning Frozen Parameters $\delta(\rho)$ via Basis Functions}
\label{sec:ILCBF}
In this section, feedforward parameters are learned in a trial-to-trial fashion, which will serve as training data for the GP, leading to contribution C2. Here, ILCBF is used to learn the parameters, but the framework can directly be extended to other feedforward parameter tuning approaches. \par
ILCBF parametrizes the feedforward signal and learns the feedforward parameters in a trial-to-trial fashion. The optimization criterion in ILCBF is specified as \cite{Bolder2014}
\begin{equation}
	\label{eq:perfCritILC}
	\footnotesize
	V\left({\theta}_{j+1}\right)=\left\|e_{j+1}(k)\right\|_{W_{e}}^{2}\!\!+\left\|f_{j+1}(k)\right\|_{W_{f}}^{2}\!\!+\left\|f_{j+1}(k)-f_{j}(k)\right\|_{W_{\Delta f}}^{2},
\end{equation}
with weighting matrices $W_e \succ 0$ and $W_f$, $W_{\Delta f}\succeq 0$ and $\theta_j\in\mathbb{R}^{n_\theta}$ the feedforward parameters. The error in trial $j+1$ can be written as
\begin{equation}
	\label{eq:errorj+1_2}
	\begin{split}
		e_{j+1}(k) &= S(q^{-1})r(k)-S(q^{-1})G_0(q^{-1}) f_{j+1}(k)\\
		&= e_j(k) - S(q^{-1})G_0(q^{-1})\Big(f_{j+1}(k)-f_j(k)\Big),
	\end{split}
\end{equation}
where now, $G_0$ is for instance a nominal model of a position-dependent system. The feedforward force is parameterized in terms of the feedforward parameters $\theta_j$, i.e., $f_j(k)=F(\theta_j)r(k)$, with $F(\theta_j) \in \mathbb{R}^{N\times N}$ the convolution matrix of a linear system with parameters $\theta_j$. ILCBF updates the feedforward parameters in a trial-to-trial fashion using
\begin{equation}
	\label{eq:feedforwardUpdate}
	{\theta}_{j+1}^* = \arg \min_{{\theta}_{j+1}} V\left({\theta}_{j+1}\right).
\end{equation}
When choosing $F(\theta_j)$ linearly in $\theta_j$, the optimization criterion in \eqref{eq:perfCritILC} becomes quadratic in $\theta_{j+1}$. Hence, an analytic solution to \eqref{eq:feedforwardUpdate} exists \cite{Bolder2015}. \par
Given the basis function matrix $\Psi(k)=\partial/\partial {\theta}_jF(\theta_j)r(k) \in \mathbb{R}^{N\times n_\theta}$ and the weighting matrices $W_e$, $W_f$ and $W_{\Delta f}$, the analytic solution to \eqref{eq:feedforwardUpdate} is
\begin{equation}
	\label{eq:feedforwardParameterUpdate}
	\begin{split}
		{\theta}_{j+1} &= Le_j+Q{\theta}_j, \\
		L&=R^{-1}\left(\Psi^{\top}G_0^{\top} S^{\top} W_{e}\right), \\
		Q&=R^{-1} \Psi^{\top}\left(G_0^{\top} S^{\top} W_{e} G_0 S+W_{\Delta f}\right) \Psi, \\
		R &= \left(\Psi^{\top}\left(G_0^{\top} S^{\top} W_{e} G_0 S+W_{f}+W_{\Delta f}\right) \Psi\right),
	\end{split}
\end{equation}
where $(q^{-1})$ and $(k)$ have been left out for brevity. The parameter update in \eqref{eq:feedforwardParameterUpdate} leads to monotonic convergence of $\|f_j(k)\|$, provided matrices $W_e$, $W_f$ and $W_{\Delta f}$ are selected properly \cite{Bolder2014}. Robustness, with respect to model mismatch due to the position-dependent dynamics, can be enforced by increasing $W_f$. Now, \eqref{eq:feedforwardParameterUpdate} can be used in combination with the error in trial $j$ to compute a new set of feedforward parameters $\theta_{j+1}$ for a fixed value of $\rho$. 
%%%%%%%%%%%%%%%%%%%%%%%%%%%%%%%%%%%%%%%%%%%%%%%%%Example%%%%%%%%%%%%%%%%%%%%%%%%%%%%%%%%%%%%%%%%%%%%%%%%%%
\section{Flexible Beam example}
\label{sec:results}
In this section, the GP snap framework is applied on a simulation of an flexible beam, hence constituting contribution C3. First, the example setup is discussed, followed by the application of both ILCBF and GP snap feedforward. Finally, the results and comparison to position-independent snap feedforward are given.
%---------------------------------------------------------------
\subsection{Example System}
The unsupported (free-free) flexible beam in Fig.~\ref{fig:beamFigure} is considered, having dominant flexible dynamics. The beam has a total length of 500 mm. The flexible beam consists of two actuators and a sensor with variable position. To assure the system is SISO, the two actuators have the same input, i.e., $a_1=a_2=u/2$. Varying performance location, e.g., as is occurring in wafer exposure \cite{Voorhoeve2021} or flat-bed printing \cite{deRozario2017}, can be emulated by changing the sensor position $\rho$. The translational rigid body and two flexible modes are considered, i.e., $n_{RB}=1$ and $n_f=2$. The feedback controller $C$ is taken fixed and is a lead filter and a gain with a bandwidth of 4 Hz. Due to the inherent dynamics of the flexible beam and a changing performance location, it is a suitable example for position-dependent snap feedforward.
%-----------------------------------------------------------------
\subsection{Learning Frozen Parameters on the Flexible Beam}
ILCBF as specified in Section~\ref{sec:ILCBF} has been implemented on the flexible beam to iteratively learn the feedforward parameters, including the snap feedforward parameter. A fourth-order reference as designed in \cite{Lambrechts2005} has been used. The model $G_0$ for ILCBF for all positions, see e.g. \eqref{eq:feedforwardParameterUpdate}, is a model of the flexible beam with sensor position $\rho=250$ mm, which is seen in Fig.~\ref{fig:beamBodes}. 
\begin{figure}[tbp]
\definecolor{mycolor1}{rgb}{0.00000,0.44700,0.74100}%
\definecolor{mycolor2}{rgb}{0.75294,0.75294,0.75294}%
\begin{tikzpicture}

\begin{axis}[%
width=0.32\columnwidth,
height = 0.5\columnwidth,
scale only axis,
xmin=0.7,
xmax=6.3,
xtick={1, 2, 3, 4, 5, 6},
xlabel style={font=\color{white!15!black}},
xlabel={Trial Number [-]},
ymin=1e-05,
ymax=0.001,
yminorticks=true,
ylabel style={font=\color{white!15!black}},
ylabel={$\|e\|_2$ [$m$]},
axis background/.style={fill=white},
ymode = log,
]
\addplot [color=mycolor1, dashed, line width=1.3pt, mark size=4pt, mark=square, mark options={solid, mycolor1}, forget plot]
  table[row sep=crcr]{%
1	0.000669602442765315\\
2	1.83194496727306e-05\\
3	1.81872789904187e-05\\
4	1.81829468609167e-05\\
5	1.81829445118532e-05\\
6	1.81829444658141e-05\\
7	1.81829444647581e-05\\
};
\addplot [color=mycolor2, dashed, line width=1.3pt, mark size=4pt, mark=o, mark options={solid, mycolor2}, forget plot]
  table[row sep=crcr]{%
1	0.000669602442765315\\
2	4.05669635286743e-05\\
3	4.05275795314416e-05\\
4	4.05275792728227e-05\\
5	4.05275792721162e-05\\
6	4.05275792721194e-05\\
7	4.05275792721184e-05\\
};
\end{axis}

\end{tikzpicture}%
\definecolor{mycolor1}{rgb}{0.00000,0.44700,0.74100}%
\definecolor{mycolor2}{rgb}{0.85000,0.32500,0.09800}%
\begin{tikzpicture}
\pgfplotsset{
    scale only axis,
    xmin=0.7, xmax=6.3,
    xlabel = Trial Number [-],
    width=0.32\columnwidth,
    xtick={1, 2, 3, 4, 5, 6},
    height = 0.5\columnwidth
}
\begin{axis}[
  axis y line*=left,
  ymin=-0.007, ymax=0.17,
  ylabel=y-axis 1,
  ylabel style={font=\color{mycolor1}},
  ylabel={Acceleration Parameter [$kg$]},
  every outer y axis line/.append style={mycolor1},
  every y tick label/.append style={font=\color{mycolor1}},
  every y tick/.append style={mycolor1},
  y tick label style={/pgf/number format/.cd,fixed,precision=2}
]
\addplot [color=mycolor1, dashed, line width=1.3pt, mark size=4.0pt, mark=triangle, mark options={solid, mycolor1}]
  table[row sep=crcr]{%
1	0\\
2	0.155584829241907\\
3	0.155046775523325\\
4	0.155050306967894\\
5	0.155050477277499\\
6	0.155050476468493\\
};
\end{axis}
\begin{axis}[
  axis y line*=right,
  axis x line=none,
  ymin=-1.3e-6, ymax=3.5e-5,
  ylabel=y-axis 2,
  ylabel style={font=\color{mycolor2}},
    ylabel={Snap Parameter [$kg/s^2$]},
    every outer y axis line/.append style={mycolor2},
every y tick label/.append style={font=\color{mycolor2}},
every y tick/.append style={mycolor2},
]
\addplot [color=mycolor2, dashed, line width=1.3pt, mark size=4.0pt, mark=o, mark options={solid, mycolor2}, forget plot]
  table[row sep=crcr]{%
1	0\\
2	2.92029760001474e-05\\
3	2.86111390904854e-05\\
4	2.86009008129915e-05\\
5	2.8601148827242e-05\\
6	2.86011534502371e-05 \\
};
\end{axis}

\end{tikzpicture}%
	\caption{Left: error 2-norm for acceleration feedforward (\protect\silverdashcircle) and acceleration with snap feedforward (\protect\bluedashsquare) when applying ILCBF. Right: acceleration (\protect\bluedashtriangle) and snap (\protect\redcircle) feedforward parameters when applying ILCBF. The scheduling variable $\rho$ is equal to 250 mm.}
	\label{fig:beamILCBFConvergence}
\end{figure}
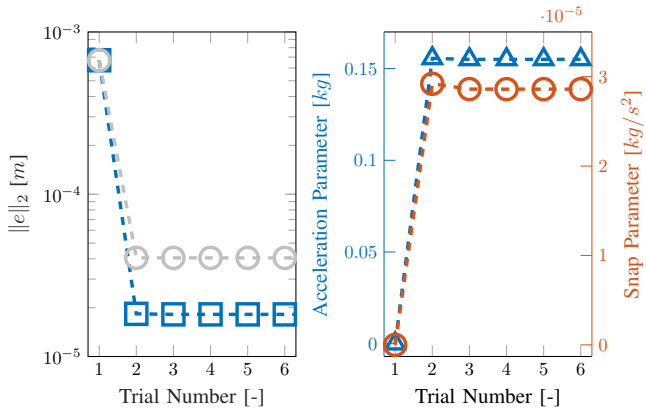
Fig.~\ref{fig:beamILCBFConvergence} shows that the error 2-norm can be significantly reduced by learning the feedforward parameters in a trial-to-trial fashion. It furthermore shows how additional snap feedforward can improve performance compared with traditional acceleration feedforward.
%---------------------------------------------------------------
\subsection{GP Snap Feedforward Parameter}
\begin{figure}[tbp]
	\vspace{0.65mm}
	\centering
	\input{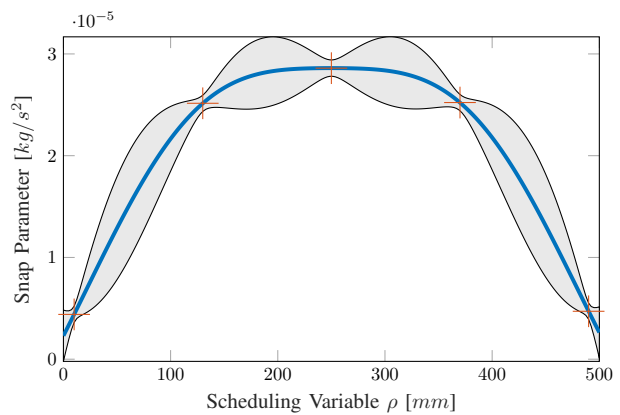}
	\caption{GP regression of the snap feedforward parameter of the flexible beam with the posterior mean (\protect\blueline), the posterior mean $\pm 2 \sigma$ (\protect\varianceband) and the training data (\protect\redplus).}
	\label{fig:GPregressionSnap}
\end{figure}

The snap feedforward parameter is modeled as a function of position using a GP. First, the training inputs in \eqref{eq:GPinputs} are defined as five equispaced positions,
\begin{equation}
	\label{eq:exampleTrainingPositions}
	P = \begin{bmatrix}10 & 130 & 250 & 370 & 490 \end{bmatrix}^\top. 
\end{equation}
On the positions $P$, ILCBF has been performed and the resulting snap feedforward parameter is the training data as in \eqref{eq:trainingData}. The RBF covariance function in \eqref{eq:RBFKernel} is used, with hyperparameters optimized based on marginal likelihood. For visualization purposes, the test positions $P_*$ are defined as an equispaced dense grid covering the beam. The GP regression in Fig.~\ref{fig:GPregressionSnap} shows the snap feedforward parameter varies when changing the scheduling variable $\rho$, which is further supported by looking at Fig.~\ref{fig:beamBodes}.
%-----------------------------------------------------------------------------------------------
\subsection{Results}
\begin{figure}[tbp]
	\centering
\definecolor{mycolor1}{rgb}{0.00000,0.44700,0.74100}%
\definecolor{mycolor2}{rgb}{0.85000,0.32500,0.09800}%
\definecolor{mycolor3}{rgb}{0.92900,0.69400,0.12500}%
\begin{tikzpicture}

\begin{axis}[%
width=0.85\columnwidth,
height = 0.47\columnwidth,
scale only axis,
bar shift auto,
xmin=0.511111111111111,
xmax=5.48888888888889,
xtick={1,2,3,4,5},
xticklabels={{30},{110},{248},{387},{470}},
xlabel style={font=\color{white!15!black}},
xlabel={Scheduling Variable ${\rho}$ [$mm$]},
ymin=0,
ymax=4.5e-05,
ylabel style={font=\color{white!15!black}},
ylabel={$\|e\|_2$ [$m$]},
]
\addplot[ybar, bar width=0.178, fill=mycolor1, draw=black, area legend] table[row sep=crcr] {%
1	8.90896909237811e-06\\
2	9.2852125295335e-06\\
3	1.81791206567016e-05\\
4	9.57451036719776e-06\\
5	8.88236354731433e-06\\
};
\addplot[forget plot, color=white!15!black] table[row sep=crcr] {%
0.511111111111111	0\\
5.48888888888889	0\\
};
\addplot[ybar, bar width=0.178, fill=mycolor2, draw=black, area legend] table[row sep=crcr] {%
1	2.05051945846605e-05\\
2	1.10100618000015e-05\\
3	1.81791856271926e-05\\
4	1.10233950957136e-05\\
5	2.03178990401297e-05\\
};
\addplot[forget plot, color=white!15!black] table[row sep=crcr] {%
0.511111111111111	0\\
5.48888888888889	0\\
};
\addplot[ybar, bar width=0.178, fill=mycolor3, draw=black, area legend] table[row sep=crcr] {%
1	9.38468225207196e-06\\
2	2.57070417644424e-05\\
3	4.29325120748369e-05\\
4	2.65827877290049e-05\\
5	9.48488147638014e-06\\
};
\addplot[forget plot, color=white!15!black] table[row sep=crcr] {%
0.511111111111111	0\\
5.48888888888889	0\\
};
\end{axis}
\end{tikzpicture}%
	\caption{Error 2-norm for GP snap (\protect\fillblue), position-independent snap (\protect\fillred) and acceleration feedforward (\protect\fillyel). The GP feedforward has comparable performance for all positions, while both the position-independent snap and acceleration feedforward have higher error 2-norms for certain positions.}
	\label{fig:error2norm}
\end{figure}
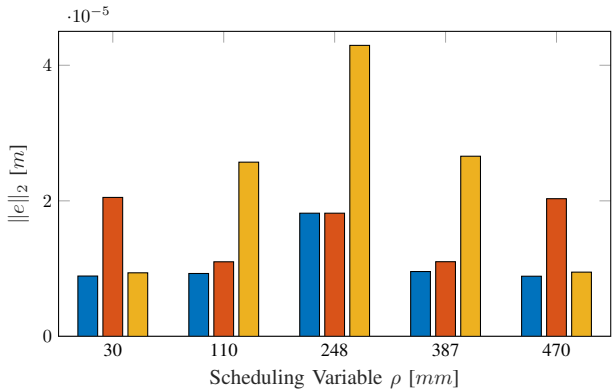

\begin{figure}[tbp]
	\centering
	\input{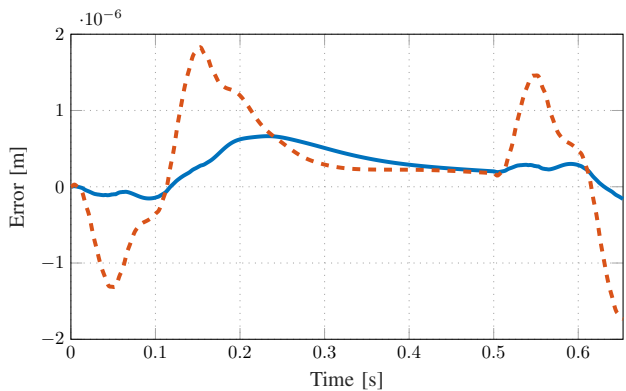}
	\caption{Time domain error for GP snap feedforward (\protect\blueline) and position-independent snap feedforward (\protect\redline) for the scheduling variable $\rho=30$ mm. The maximum error for GP snap feedforward is roughly 3 times lower than the maximum error for position-independent snap feedforward.}
	\label{fig:errorTime}
\end{figure}

To evaluate the performance of the framework, GP snap feedforward is compared with position-independent snap feedforward. Position-independent snap feedforward uses the same snap feedforward parameters for all values of $\rho$, whereas the GP snap feedforward estimates the snap feedforward parameter using \eqref{eq:estTheta}. The error 2-norm for both GP snap and position-independent snap feedforward for several test positions can be seen in Fig.~\ref{fig:error2norm}.
Fig.~\ref{fig:error2norm} shows that GP snap feedforward outperforms the position-independent feedforward significantly when the sensor position moves outside the center position. At the edges of the flexible beam, roughly a performance increase of factor two in terms of the error 2-norm is observed. Near the center of the beam, performance is equal, which is expected since the position-independent snap feedforward uses the snap feedforward parameter determined at the center position. The achievable performance gain is further supported by looking at the time domain error for $\rho=30$ mm in Fig.~\ref{fig:errorTime}.

%%%%%%%%%%%%%%%%%%%%%%%%%%%%%%%%%%%%%%%%%%%%%%%%%Conclusion%%%%%%%%%%%%%%%%%%%%%%%%%%%%%%%%%%%%%%%%%%%%%%%
\section{Conclusions}
\label{sec:conclusions}
This work describes a method to use a GP to model the snap feedforward parameter as a function of position. A GP works especially well since it is non-parametric and therefore does not assume a parametric form between the position and the snap feedforward parameter, which is typically unknown. The framework is applied on a flexible beam, which shows the dependency of the snap feedforward parameter on position. GP snap feedforward shows a significant performance increase compared with position-independent snap feedforward. \par
Future research on this topic is directed at integrating and testing the framework for MIMO systems and adding other position-dependent feedforward parameters or effects. Furthermore, a method to automatically and optimally determine the training inputs in \eqref{eq:exampleTrainingPositions} is investigated. Lastly, experimental validation confirming the practical applicability of the framework is a subject of ongoing research.
%%%%%%%%%%%%%%%%%%%%%%%%%%%%%%%%%%%%%%%%%%%%%%%%%Ackowledgment%%%%%%%%%%%%%%%%%%%%%%%%%%%%%%%%%%%%%%%%%%%%
%\section*{Acknowledgment}
%The authors would like to thank Jim Portegies, Dragan Kosti\'c and Robin van Es for their contributions to this research. Furthermore, the support from ASM Pacific Technology is greatly appreciated.

%-------------------------------------------bibliography---------------------------------------------------

\bibliographystyle{IEEEtran}
\bibliography{IEEEabrv,references}

\end{document}